\documentclass[final,5p,times,twocolumn]{elsarticle}

\usepackage{lineno,hyperref}
\modulolinenumbers[5]

\usepackage{siunitx}
\usepackage{textcomp} 
\DeclareSIUnit\permille{\text{\textperthousand}}
\journal{Proceedings of the 15th Pisa Meeting on Advanced Detectors}

\begin{document}

\begin{frontmatter}

\title{Towards a New Generation of Monolithic Active Pixel Sensors}

\author[]{Ankur Chauhan}
\author[]{Manuel Del Rio Viera\fnref{bonn}}
\author[]{Doris Eckstein}
\author[]{Finn Feindt\corref{corresAuthor}}
	\cortext[corresAuthor]{Corresponding author}
	\ead{finn.feindt@desy.de}
\author[]{Ingrid-Maria Gregor\fnref{bonn}}
\author[]{Karsten Hansen}
\author[]{Lennart Huth}
\author[]{Larissa Mendes\fnref{camp}}
\author[]{Budi Mulyanto}
\author[]{Daniil Rastorguev\fnref{wupp}}
\author[]{Christian Reckleben}
\author[]{Sara Ruiz Daza\fnref{bonn}}
\author[]{Paul Schütze}
\author[]{Adriana Simancas\fnref{bonn}}
\author[]{Simon Spannagel}
\author[]{Marcel Stanitzki}
\author[]{Anastasiia Velyka}
\author[]{Gianpiero Vignola\fnref{bonn}}
\author[]{Håkan Wennlöf}
\address{Deutsches Elektronen-Synchrotron DESY, Notkestr. 85, 22607 Hamburg, Germany}

\fntext[bonn]{Also at University of Bonn, Germany}
\fntext[camp]{Also at University of Campinas, Brazil}
\fntext[wupp]{Also at University of Wuppertal, German}

\begin{abstract}
A new generation of Monolithic Active Pixel Sensors (MAPS), produced in a \SI{65}{\nano\meter}
CMOS imaging process, promises higher densities of on-chip circuits and, for a given pixel 
size, more sophisticated in-pixel logic compared to larger feature size processes. MAPS are a
cost-effective alternative to hybrid pixel sensors since flip-chip bonding is not required. In
addition, they allow for significant reductions of the material budget of detector systems,
due to the smaller physical thicknesses of the active sensor and the absence of a separate
readout chip.

The TANGERINE project develops a sensor suitable for future Higgs factories as well as for a beam
telescope to be used at beam-test facilities. The sensors will have small collection electrodes 
(order of \SI{}{\micro\meter}) to maximize the signal-to-noise ratio, which makes it possible to minimize 
power dissipation in the circuitry. 
The first batch of test chips, featuring full front-end amplifiers with Krummenacher
feedback, was produced and tested at the Mainzer Mikrotron (MAMI) at the end of 2021. MAMI
provides an electron beam with currents up to \SI{100}{\micro\ampere} and an energy of 
\SI{855}{\mega\electronvolt}. The analog output signal of the test chips was recorded with a high
bandwidth oscilloscope and used to study the charge-sensitive amplifier of the chips in terms of
waveform analysis. A beam telescope was used as a reference system to allow for track-based 
analysis of the recorded data.
\end{abstract}


\end{frontmatter}


	\section{Introduction}
The Tangerine (Towards Next Generation Silicon Detectors) project pursues the goal of developing a monolithic active pixel sensor (MAPS) using a \SI{65}{\nano\meter} CMOS imaging process. The sensor will be optimized for the requirements of e.g. future lepton colliders, so the project aims for a spatial resolution below \SI{3}{\micro\meter},
temporal resolution below \SI{10}{\nano\second}, and a total thickness below \SI{50}{\micro\meter}.

To optimize the sensor design, the project employs a chain of simulation tools predicting the performance of a specific sensor design in a tracking application. The first step is the detailed calculation of the electric fields, starting from a generic doping profile. To do so, the Poisson equations are numerically solved, using Synopsys Technology Computer Aided Design (TCAD) software~\cite{sentaurusTCAD}. These electric fields are used in Allpix$^2$~\cite{ap2} for charge transport simulations. The energy deposition by charged particles is simulated via an interface to Geant4~\cite{geant4}. 

The simulated performance is compared for different sensor designs, emphasizing sensor volume modifications as introduced in~\cite{modifiedProcess,ngapProcess} for sensors produced in a \SI{180}{\nano\meter} CMOS imaging process. Also, the pixel pitch, the biasing conditions, the width of the p-well opening and the width of the gap in the n-blanket are varied. More details on the sensor layout and the simulation procedure are given in~\cite{hw_tangerine}.

This paper addresses the characterization of a first test chip featuring Krummenacher type charge sensitive amplifiers (CSA)~\cite{krummenacher}, received in October 2021. The main feature of the amplifier is a continuous reset, well suited for time over threshold (TOT) measurements. The CSA test chip features two CSAs with different feedback capacitances (1.5 and \SI{2}{\femto\farad}), which can be investigated via test-pulse injection. It also hosts a $2 \times 2$ pixel matrix with a pitch of \SI{16.3}{\micro\meter}. The output signal of these pixels is amplified with the same type of CSA (\SI{2}{\femto\farad}) and an additional operational amplifier before they are recorded with a high-bandwidth oscilloscope (\SI{4}{\giga\hertz}, \SI{10}{\giga s \per\second} per channel) in edge-trigger mode.

	\section{Sensor Testing}
First tests of the CSA test chip were performed in the laboratory using an $^{55}$Fe source. They were followed by a campaign of test-beam measurements at the DESY II Test Beam facility~\cite{desyII}, CERN SPS, and at the Mainzer Mikrotron (MAMI) facility~\cite{mamiB}, using a EUDET-type~\cite{eudet}, a Timepix-based~\cite{timepix3}, and a compact ALPIDE-based~\cite{alpide} beam telescope, respectively. The acquired waveforms were analyzed in terms of a pulse shape analysis. For an analysis including track reconstruction, the Corryvreckan software~\cite{curry} was used. Data were recorded with trigger rates on the order of \SI{0.001}{\hertz}, \SI{0.01}{\hertz} and \SI{1}{\hertz} at DESY, SPS and MAMI, respectively. This is about two orders of magnitude lower than expected for the active area of the chip and attributed to an issue in the sensor design, leading to low detection efficiencies everywhere but under the read-out electrodes.

\begin{figure}[tbp]
	\centering
	\includegraphics[width=0.5\textwidth]{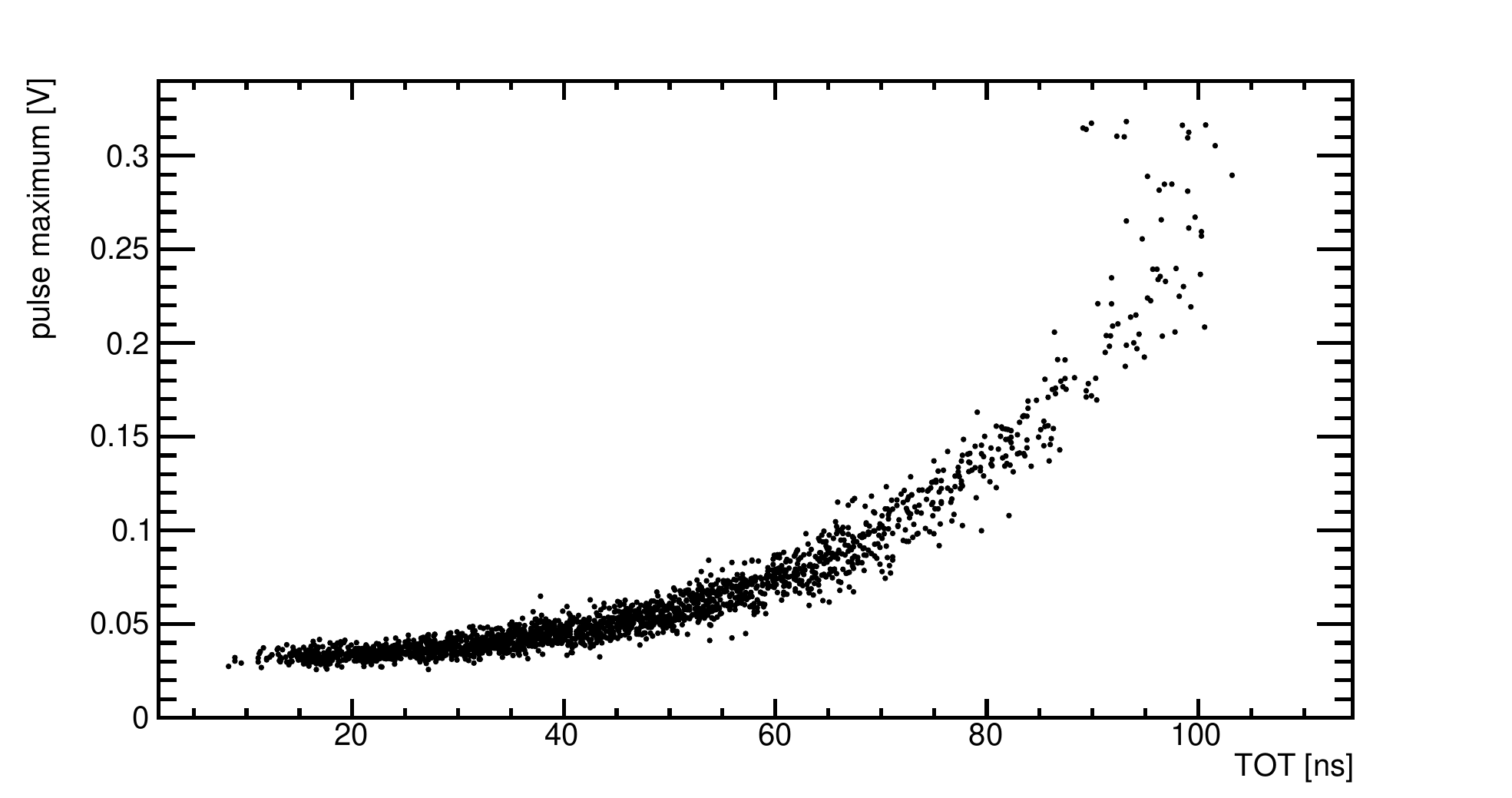}
	\caption[]{Baseline-corrected maximum of the amplified pulses as a function of TOT, measured with a CSA test chip at the MAMI facility. The applied threshold corresponds to about \SI{27}{\milli\volt}.}
	\label{fig::ampVsTot}
\end{figure}

MAMI provides a beam of electrons with an energy of \SI{855}{\mega\electronvolt} and beam currents up to \SI{100}{\micro\ampere}. The waveforms recorded at MAMI were analyzed and figure~\ref{fig::ampVsTot} shows the maximum of the pulse, after subtraction of the baseline, as a function of the TOT. The small slope for $\text{TOT} <$ \SI{50}{\nano\second} is not desired, as a small slope makes the TOT less sensitive to signal variations in this region, which contains the signal expected for a minimum ionizing particle. Improvements are foreseen for the next version of the front-end amplifier by reducing the circuit's slew-rate dependence on the amplitude. The rise time of the output pulses is limited by the operational amplifiers. It was found to be in the range of 4 to \SI{9}{\nano\second}, depending on the investigated sample and the operational parameters of the CSA test chip.

\begin{figure}[tbp]
	\centering
	\includegraphics[width=0.5\textwidth]{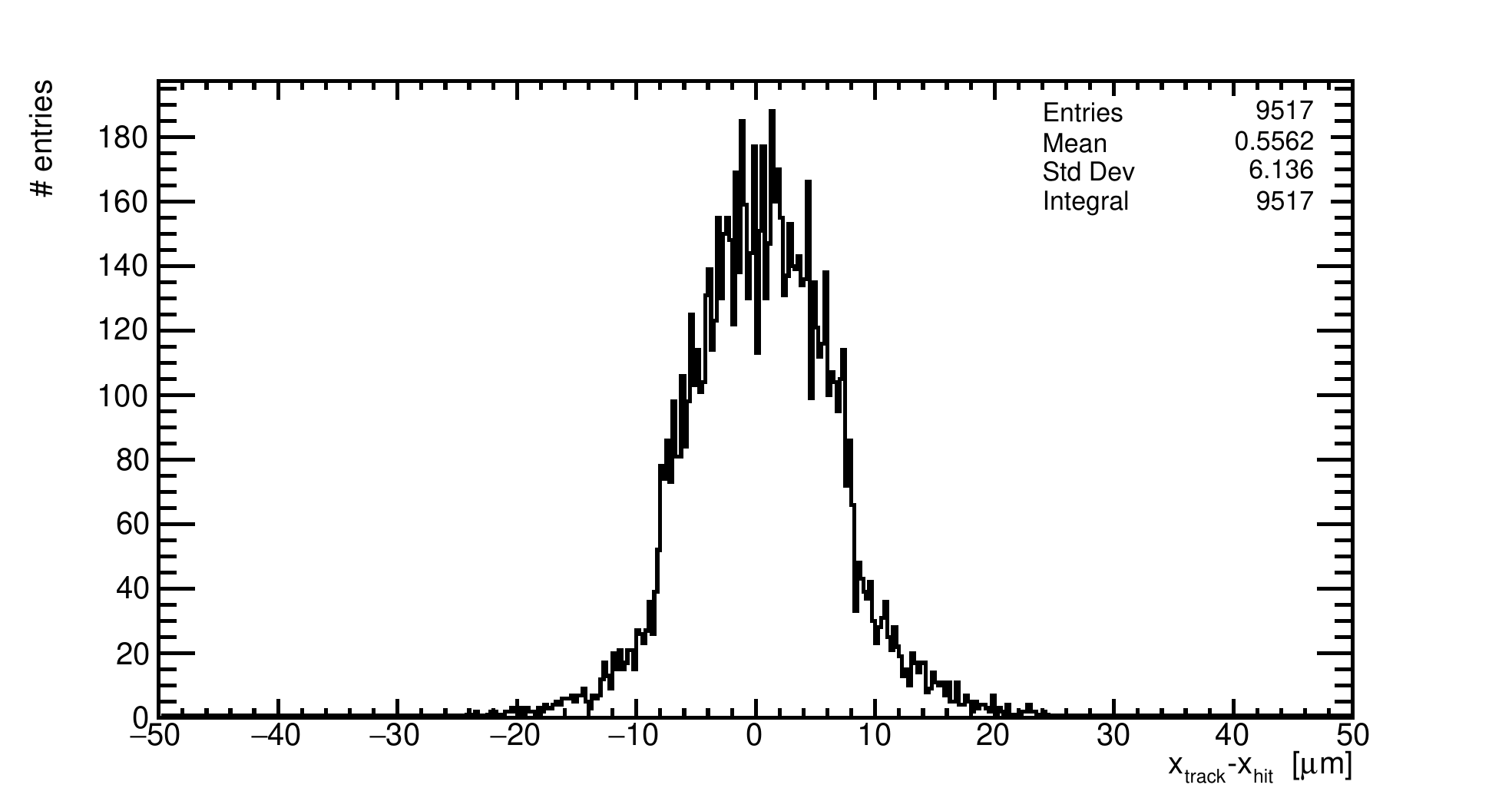} 
	\caption[]{Residual distribution between the position measurement of the CSA test chip and the reference track reconstructed with the beam telescope. The measurements were carried out at the MAMI facility.}
	\label{fig::residual}
\end{figure}

The track-based analysis of the data acquired at CERN delivered further evidence for the low efficiencies mentioned above. Despite this, the data taken at MAMI made it possible to reconstruct residuals between the intersection of the reconstructed electron track with the CSA test chip ($x_{track}$), and the chips own position measurements ($x_{hit}$), presented in figure~\ref{fig::residual}. The residual width (Std Dev) is dominated by the interpolation error of the beam telescope, which is on the order of \SI{5}{\micro\meter} for the \SI{855}{\mega\electronvolt} electrons and the spacing of \SI{20}{\milli\meter} between all adjacent planes. It should be mentioned that the position resolution of the CSA test chip is artificially enhanced by the small sensitive area. Still, this figure proves that the test chip is successfully integrated with the reference system and works as a detector for charged particles.

	\section{Conclusion and Outlook}
These results present the first successful operation of \SI{65}{\nano\meter} CMOS sensors developed at DESY. This sensor is dedicated to a first characterization of the new CSA design, which will be improved in terms of TOT linearity for future sensors. The rise time of the output pulses of the CSA is found to be in the range of 4 to \SI{9}{\nano\second}, limited by the operational amplifiers, so that time resolutions on the order of \SI{}{\nano\second} are expected to be achievable. A flaw in the pixel design, leading to an efficient region of only a few \SI{}{\micro\meter} around the readout electrode, has been identified and fixed for future submissions.

A second generation of test chips is expected for the beginning of 2023 and will include an upgraded version of the CSA test chip and a fully integrated chip with $64 \times 16$ square pixels of \SI{35}{\micro\meter} pitch and an \SI{8}{bit} counter per pixel. In the meantime a set of Analogue Pixel Test Structures (APTS)~\cite{APTS1,APTS2}, featuring pitches of 15, 20 and \SI{25}{\micro\meter} and the standard, n-blanket and n-gap design, will be characterized. The results will be used to validate the simulation procedure and improve its predictive power for the optimization of the final TANGERINE sensor design.

	\section{Acknowledgments}
The Tangerine project receives funding from the Helmholtz Innovation Pool, 2021–2023. Measurements leading to some of the presented results have been performed at the Test Beam Facility at DESY Hamburg (Germany), a member of the Helmholtz Association (HGF). Some of the developments presented in this contribution are performed in collaboration with the CERN EP R\&D programme on technologies for future experiments. 

The authors wish to express their gratitude to the Institut f\"ur Kernphysik at the Johannes Gutenberg-University in Mainz, and the MAMI test beam and its operating crew. We are also grateful for the help and support from the ALICE ITS3 project, and the CERN SPS test beam facility.	
\bibliography{mybibfile}

\begin{thebibliography}{10}
\expandafter\ifx\csname url\endcsname\relax
  \def\url#1{\texttt{#1}}\fi
\expandafter\ifx\csname urlprefix\endcsname\relax\def\urlprefix{URL }\fi
\expandafter\ifx\csname href\endcsname\relax
  \def\href#1#2{#2} \def\path#1{#1}\fi

\bibitem{sentaurusTCAD}
Synopsys, \href{https://www.synopsys.com/silicon/tcad.html}{{TCAD} -
  {Technology} {Computer} {Aided} {Design}}, accessed 2022-02-18 (2022).
\newline\urlprefix\url{https://www.synopsys.com/silicon/tcad.html}

\bibitem{ap2}
S.~Spannagel, K.~Wolters, D.~Hynds, et~al., Allpix$^2$: A modular simulation
  framework for silicon detectors, Nucl. Instrum. Methods Phys. Res. A 901
  (2018) 164--172.
\newblock \href {https://doi.org/10.1016/j.nima.2018.06.020}
  {\path{doi:10.1016/j.nima.2018.06.020}}.

\bibitem{geant4}
S.~Agostinelli, J.~Allison, K.~Amako, et~al., Geant4—a simulation toolkit,
  Nucl. Instrum. Methods Phys. Res. A 506~(3) (2003) 250--303.
\newblock \href {https://doi.org/10.1016/S0168-9002(03)01368-8}
  {\path{doi:10.1016/S0168-9002(03)01368-8}}.

\bibitem{modifiedProcess}
W.~Snoeys, G.~{Aglieri Rinella}, H.~Hillemanns, et~al., A process modification
  for {CMOS} monolithic active pixel sensors for enhanced depletion, timing
  performance and radiation tolerance, Nucl. Instrum. Methods Phys. Res. A 871
  (2017) 90--96.
\newblock \href {https://doi.org/10.1016/j.nima.2017.07.046}
  {\path{doi:10.1016/j.nima.2017.07.046}}.

\bibitem{ngapProcess}
M.~Munker, M.~Benoit, D.~Dannheim, et~al.,
  \href{https://doi.org/10.1088/1748-0221/14/05/c05013}{Simulations of {CMOS}
  pixel sensors with a small collection electrode, improved for a faster charge
  collection and increased radiation tolerance}, Journal of Instrumentation
  14~(05) (2019) C05013--C05013.
\newblock \href {https://doi.org/10.1088/1748-0221/14/05/c05013}
  {\path{doi:10.1088/1748-0221/14/05/c05013}}.
\newline\urlprefix\url{https://doi.org/10.1088/1748-0221/14/05/c05013}

\bibitem{hw_tangerine}
H.~Wennlöf, A.~Chauhan, M.~D.~R. Viera, et~al., The tangerine project:
  Development of high-resolution 65 nm silicon maps, Nucl. Instrum. Methods
  Phys. Res. A (2022) 167025\href {https://doi.org/10.1016/j.nima.2022.167025}
  {\path{doi:10.1016/j.nima.2022.167025}}.

\bibitem{krummenacher}
F.~Krummenacher, Pixel detectors with local intelligence: an {IC} designer
  point of view, Nucl. Instrum. Methods Phys. Res. A 305~(3) (1991) 527--532.
\newblock \href {https://doi.org/10.1016/0168-9002(91)90152-G}
  {\path{doi:10.1016/0168-9002(91)90152-G}}.

\bibitem{desyII}
R.~Diener, J.~Dreyling-Eschweiler, H.~Ehrlichmann, et~al., The {DESY II} test
  beam facility, Nucl. Instrum. Methods Phys. Res. A 922 (2019) 265--286.
\newblock \href {https://doi.org/10.1016/j.nima.2018.11.133}
  {\path{doi:10.1016/j.nima.2018.11.133}}.

\bibitem{mamiB}
T.~Walcher, The {Mainz} microtron facility {MAMI}, Prog. Part. Nucl. Phys. 24
  (1990) 189--203.
\newblock \href {https://doi.org/10.1016/0146-6410(90)90016-W}
  {\path{doi:10.1016/0146-6410(90)90016-W}}.

\bibitem{eudet}
H.~Jansen, S.~Spannagel, J.~Behr, et~al., {Performance of the EUDET-type beam
  telescopes}, EPJ Techniques and Instrumentation 3~(7) (2006) 1--20.
\newblock \href {https://doi.org/10.1140/epjti/s40485-016-0033-2}
  {\path{doi:10.1140/epjti/s40485-016-0033-2}}.

\bibitem{timepix3}
T.~Poikela, J.~Plosila, T.~Westerlund, et~al.,
  \href{https://doi.org/10.1088/1748-0221/9/05/c05013}{Timepix3: a 65k channel
  hybrid pixel readout chip with simultaneous {ToA}/{ToT} and sparse readout},
  Journal of Instrumentation 9~(05) (2014) C05013--C05013.
\newblock \href {https://doi.org/10.1088/1748-0221/9/05/c05013}
  {\path{doi:10.1088/1748-0221/9/05/c05013}}.
\newline\urlprefix\url{https://doi.org/10.1088/1748-0221/9/05/c05013}

\bibitem{alpide}
G.~Aglieri, C.~Cavicchioli, P.~L. Chalmet, et~al., Monolithic active pixel
  sensor development for the upgrade of the {ALICE} inner tracking system,
  Journal of Instrumentation 8~(12) (2013) C12041--C12041.
\newblock \href {https://doi.org/10.1088/1748-0221/8/12/c12041}
  {\path{doi:10.1088/1748-0221/8/12/c12041}}.

\bibitem{curry}
D.~Dannheim, K.~Dort, L.~Huth, et~al., Corryvreckan: a modular 4d track
  reconstruction and analysis software for test beam data, Journal of
  Instrumentation 16~(03) (2021) P03008.
\newblock \href {https://doi.org/10.1088/1748-0221/16/03/p03008}
  {\path{doi:10.1088/1748-0221/16/03/p03008}}.

\bibitem{APTS1}
A.~Kluge, \href{https://indico.cern.ch/event/1044975/}{{ALICE ITS3} -- a bent,
  wafer-scale {CMOS} detector}, presented at the 16th Vienna Conference on
  Instrumentation (Feb 2022).
\newline\urlprefix\url{https://indico.cern.ch/event/1044975/}

\bibitem{APTS2}
G.~A. Rinella, \href{https://agenda.infn.it/event/22092/}{Developments of
  stitched monolithic pixel sensors towards the application in the {ALICE
  ITS3}}, presented at the 15th Pisa Meeting on Advanced Detectors (May 2022).
\newline\urlprefix\url{https://agenda.infn.it/event/22092/}

\end{thebibliography}

\end{document}